**Title:** Becoming Good Stewards of Information: A framework for integrating ethical, civic, and professional formation throughout the statistics and data science curriculum

**Author:** Kaitlyn G Fitzgerald, Villanova University, Department of Mathematics & Statistics, Villanova, PA, 19085, Kaitlyn.fitzgerald@villanova.edu, https://orcid.org/0000-0001-6569-4494

**Abstract:** Recent recommendations in statistics and data science education emphasize goals that extend beyond content mastery, including statistical literacy, evidence-based decision-making, communication, ethics, responsible use of data, and civic responsibility. We argue these goals can all be understood through a common lens: helping students become better *stewards of information*. Stewardship of information provides a unifying framework for the formation of students as statisticians, data scientists, and citizens who are thoughtful producers and consumers of data and evidence, communicate findings with integrity and transparency, engage critically with data ethics, recognize the human beings behind the data, and harness data to make informed decisions, serve communities, and advance the common good. We present stewardship of information as a developmental model that can be introduced early and revisited with increasing sophistication across a curriculum, including in introductory, advanced, applied, and theoretical courses. To illustrate this approach, we provide concrete examples of course activities and curated weekly readings as a low-barrier implementation strategy and present preliminary evidence of student engagement with the framework. By offering both a conceptual foundation and practical implementation strategies, this paper contributes a coherent approach for integrating ethical, professional, and civic formation across the statistics and data science curriculum.

**Acknowledgements.** I would like to thank Azusa Pacific University, which provided several professional development opportunities that contributed to the ideas and curriculum development presented here.

**AI Disclosure:** The author utilized ChatGPT 5.3 and M365 Copilot GPT-5 for editorial feedback on manuscript organization, clarity, and concision. All substantive ideas and intellectual content are solely the author's.

**Disclosure Statement:** The author has no conflicts of interest to report.

**IRB:** Student quotations included in this paper are drawn from course artifacts collected under an Institutional Review Board at Azusa Pacific University (#23-141, exempt status) and/or were used with explicit student permission.

**Data Availability Statement:** The author confirms that the data supporting the findings of this study are available within the article and its supplementary material.

## 1. Introduction

Statistical literacy and evidence-based decision-making have long been emphasized as foundational learning goals in statistics courses (American Statistical Association, 2005). Increasingly, communication, ethics, responsible use of data, and civic responsibility have also been recognized as crucial to statistics and data science education (e.g., GAISE College Report ASA Revision Committee, 2016, 2025; International Association of Statistics Education, n.d.; National Academies of Sciences, Engineering, and Medicine, 2018). While many may agree these goals have merit, meaningfully integrating them into curriculum can be challenging, especially if viewed as separate learning objectives above and beyond statistical content to be covered. We find value, both philosophically and pedagogically, in considering all of statistics and data science – including the learning goals named here – to be unified under a common framework of *stewardship of information*. Stated differently, we believe statistics and data science education, fundamentally, offers a meaningful opportunity to teach and to learn how to become *better stewards of information*; that is, people

1) who are thoughtful producers and consumers of data and evidence,
2) who exhibit integrity and transparency in communication of findings,
3) who think critically about data ethics,
4) who recognize the human beings behind the data, and
5) who harness data to make informed decisions, serve communities, and advance the common good.

In this way, literacy, communication, ethics, and civic responsibility are not separate educational objectives or content to be covered but rather interconnected dimensions of the *formation* of students into thoughtful statisticians, data scientists, and citizens in the world. By formation we mean the holistic development of students' dispositions, habits of mind, and professional and civic identities.

Viewing these goals through the lens of stewardship of information offers several pedagogical advantages. First, it communicates a student-centered approach that signals we care about students as whole people, not simply as "brains on a stick" (Smith, 2016). Second, it provides a common language for articulating the "why" of statistics and data science education as a whole and a mechanism through which "extra" topics such as communication, ethics, and civic responsibility can be more naturally integrated into existing courses. Rather than treating these as separate initiatives, stewardship situates them within a coherent vision that can be introduced early and revisited throughout students' statistical development. Third, a shared framework creates opportunities for these goals to reinforce one another across a curriculum. As students encounter stewardship of information in different contexts and courses, the individual components become more than the sum of their parts, contributing to a deeper and more integrated sense of professional and civic responsibility. That is, instead of only asking "*What do we want students to learn in our courses?*" this framing points us towards deeper formational questions such as "*Who* *are we helping students become?" and "What types of statisticians, data scientists, citizens, and human beings are being formed in our classrooms?"*

We see many promising examples in the literature that implicitly adopt such a disposition, particularly those organized around themes of statistics for social good or data ethics. However, these resources are largely concentrated in introductory and applied course contexts. Less attention has been given to how these broader formational aims might be cultivated across an entire statistics and data science curriculum, including in probability, mathematical statistics, and other theoretically oriented courses.

The goals of this paper are threefold: 1) to present stewardship of information as a unifying conceptual framework for student formation in statistics and data science education, 2) to demonstrate it as a developmental model that can be revisited across courses with increasing sophistication, and 3) to provide concrete examples of weekly readings as a low-barrier implementation strategy. Importantly, our goal is not to convince readers that this is *the* proper way of understanding our field or that they must adopt explicit language of stewardship in their teaching. Rather, we hope to provoke deeper consideration of student formation and demonstrate what we have found to be a low-barrier, high-payoff approach for advancing many learning goals the statistics and data science community has deemed important but that instructors often struggle to integrate.

The paper is organized as follows. In Section 2, we expound on our conceptual definition of stewardship of information, demonstrating how many existing learning objectives can be understood through this lens. In Section 3, we briefly discuss our overarching pedagogical approach. Then, in Section 4, we provide examples for how we have utilized this framing to meaningfully engage students with ethical, civic, and professional formation throughout the curriculum, including in introductory statistics, probability, data science, and upper division courses. Section 5 provides preliminary evidence of student engagement with the theme of stewardship of information, reflection on implementation and ongoing challenges, and alignment with GAISE and other evidence-based and culturally-responsive pedagogy. We conclude in Section 6.

## 2. Stewardship of Information as a Unifying Framework

We live in the Information Age, where data abounds. While many definitions of statistics and data science could be posited, most would agree that gleaning insight from data is at the heart of our discipline. Fundamentally, we believe data and statistics can be tools to unearth knowledge, to illuminate injustices, to point towards equitable solutions, and to tell diverse stories that help us learn about the state of the world from a perspective beyond our own lived experience. Importantly, raw data cannot achieve these goals; data must be *stewarded* by humans. Stewardship is defined as *the careful and responsible management of something entrusted to one's care* (Merriam-Webster, 2026). Often, the word is used in contexts of environmental stewardship (caring for the environment, climate justice) or financial stewardship (responsibly managing personal or corporate finances). Like the environment or finances, data can be viewed as a *resource* that can be used in ethical or unethical ways and can be exploited for ill or stewarded for good.

Figure 1 provides a visual representation of stewardship of information as a unifying theme in statistics and data science education. We believe it draws together four existing goals or pedagogical emphases: statistical literacy & evidence-based decision making, statistical & data communication, data ethics & responsible use of data, and data for social good & civic statistics. Further, it provides an infrastructure through which we can emphasize two additional cross-cutting themes in our curriculum that we believe to be rich for student formation: statistics and data science as human endeavors, and uncertainty & intellectual humility. We will discuss each of these dimensions below and provide concrete examples for how to integrate them across the curriculum in Section 4.

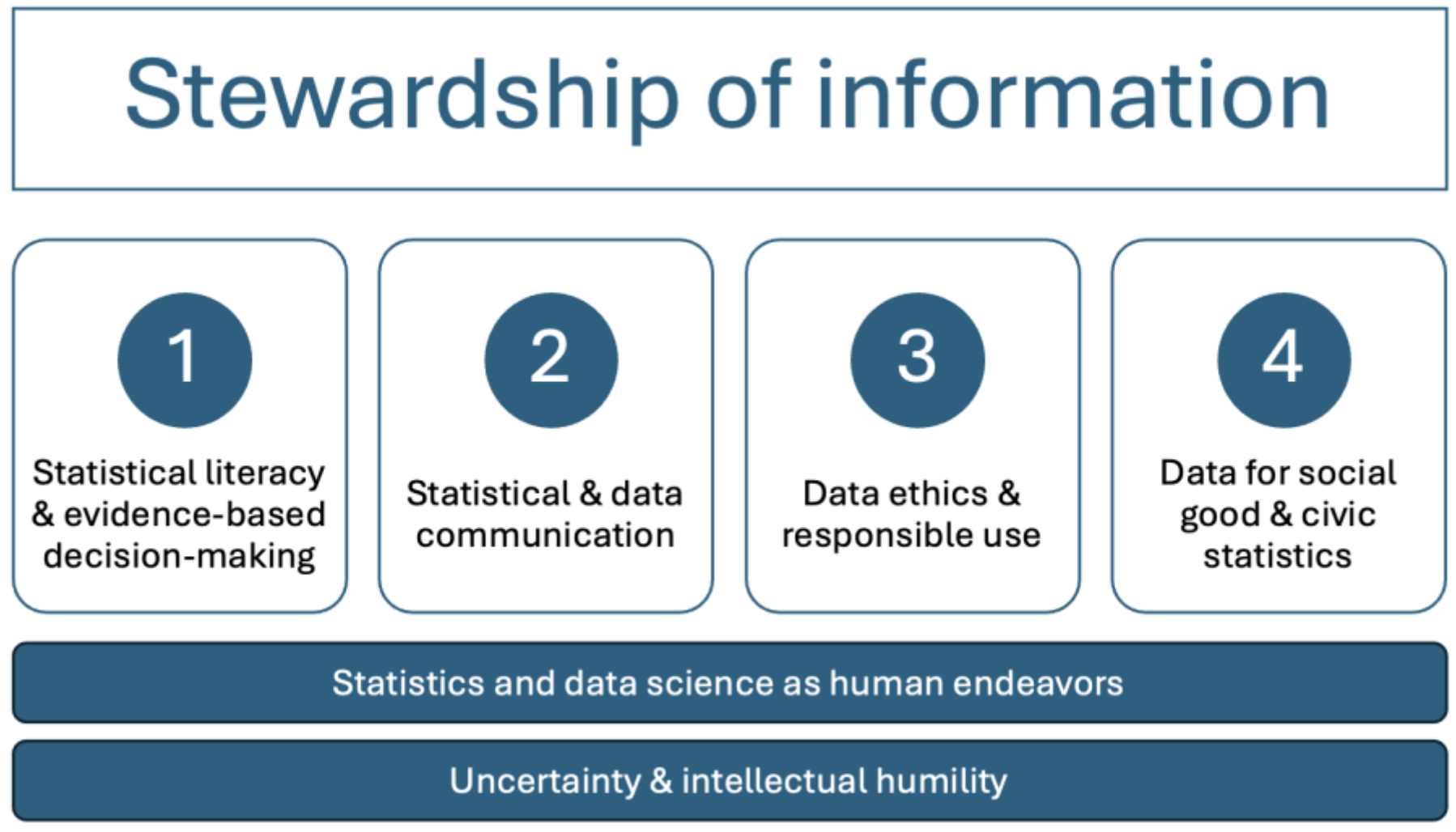

*Figure 1: Dimensions of stewardship of information*

### 2.1 Statistical Literacy & Evidence-Based Decision-Making

In the broadest sense, we consider all topics in a statistics or data science course to be connected to the common goal of developing students as good stewards of information. Even everyday technical content in our courses such as appropriately using and interpreting t-tests or regression models is a way of stewarding information well, as statistical methods are a primary means by which we are able to glean meaningful insight from data. There are also many examples of poor stewardship of information that statistics instructors routinely teach their students to avoid: p-hacking, neglecting uncertainty or variability, over-implying causation from correlation, or over-generalizing from non-random or biased samples. In this way, when our courses set out to promote statistical literacy or sound evidence-based decision-making, we are promoting good stewardship of information. While this reframing from “statistical literacy” or “evidence-based decision-making” to “stewarding information” may seem unnecessary or purely semantic, we believe it creates a larger umbrella that allows us to connect the day-to-day learning of statistical methods to the deeper ethical, civic, and professional formation that we might hope to achieve in our courses.

### 2.2 Statistical & Data Communication

If literacy is concerned with the *understanding* of statistical ideas, a natural next step of stewardship is *communicating* those ideas effectively and responsibly. As emphasized in the 2025 GAISE recommendations: “producing results in only one step in learning from data; students must also communicate their findings clearly and accurately, both orally and in writing." The guidelines go on to articulate several dimensions of what we would consider good stewardship of information in the realm of communication:

> "responsiveness to both the genre of communication and audience background, ... understanding and conveying the provenance of data, emphasizing their effect on conclusions... [presenting] inferences, visualizations, and analyses as evidence, not proof, of conclusions... and transparently acknowledging limitations and contextualizing results at an audience-appropriate technical level." (GAISE College Report ASA Revision Committee 2025, lightly condensed)

To be a thoughtful communicator of data in these ways is to be a good steward of information. Stewardship as a concept evokes an outward posture - one that points to responsibility we have to something outside ourselves. In this same way, good data communication requires an outward posture that is attuned to the needs of an audience and communities who will consume or otherwise be affected by the data and analyses.

**2.3 Data Ethics**

With rapid technological advancements and the rise of data science, many recognize the importance of ethics education, and there has been an increasingly robust emphasis on ethics in the literature (e.g., Utts 2021; Baumer et al. 2022; Raman et al. 2023). The word "ethics" or "ethical" did not appear in the 2005 GAISE guidelines, but the 2016 guidelines include a goal for introductory statistics courses pertaining to ethics in statistical practice (American Statistical Association, 2005; GAISE College Report ASA Revision Committee, 2016). While ethics cuts across the learning goals already discussed (e.g., appropriate use of methods, evidence-based decision-making, and thoughtful communication), it also introduces other important dimensions of stewardship of information, including thoughtfulness and intentionality about data privacy, transparency, consent, and bias. Both a 2018 National Academies report on undergraduate Data Science and the updated 2025 GAISE recommendations assert that ethics should be integrated throughout the statistics and data science curriculum (Canner et al., 2025; National Academies of Sciences, Engineering, and Medicine, 2018). Framing ethics as a key formational pillar of becoming good stewards of information can help facilitate its integration across courses, as will be demonstrated in Section 4. Notably, we will discuss implementation in theoretical courses such as probability and mathematical statistics where resources are more scarce.

**2.4 Data for Social Good & Civic Statistics**

While not often appearing explicitly in learning goals, "data for social good" is a framing used widely across statistics and data science education as well as industry, and we consider it to be not only a compelling pedagogical emphasis but also one that provides wide-ranging examples of what it looks like to steward data for good. Many have offered resources for teaching statistics with a social justice lens (e.g., Lesser 2007; Baumer et al. 2022; Elisa Raffaghelli 2020; Barker et al. 2025; Weiland and Williams 2024). For example, instructors might bring examples into their classroom to show students how data can be used to pursue climate justice, inform allocation of resources in anti-poverty efforts, or illuminate racial discrimination in the criminal justice system. Initiatives such as ProCivicStat through the International Association for Statistics Education represent a robust vision for the importance of statistics education in forming engaged citizens:

"We believe that students should see the usefulness of statistics in understanding evidence, and should engage with data about important social phenomena, to support their development as active and empowered citizens. Social phenomena are complex, and democracies need citizens who can explore, understand, and reason about information of a multivariate nature." (International Association of Statistics Education, n.d.)

We view this, too, as an apt articulation of a need for good stewards of information. Civic statistics and data for social good broadens the stewardship landscape to emphasize not only individual responsibility in understanding and communicating statistical ideas responsibly, but also the role that data – and good stewards of data – play in civic life and healthy democracies.

We as instructors have an opportunity to steward information in the way that we curate course content, data sets, and examples in such a way that can contribute to the holistic formation of students as engaged citizens. While we hope that emphasizing data for social good in our courses may encourage some students to pursue service-oriented careers, more broadly, we hope that it instills in students a posture of curiosity and concern for their communities and the world, irrespective of their specific career path.

**2.5 Cross-Cutting Themes**

Emphasizing statistics and data science education as a formational process of becoming good stewards of information allows us to draw out two additional themes that we believe permeate our discipline and are powerful for student formation but often remain unrecognized or under-cultivated in statistics and data science curriculum. First is the recognition that statistics and data science are ultimately human endeavors. Not only is much of the data we work with about humans and communities, but the practice of statistics and data science is done by humans and for humans. Being good stewards of information, then, requires not only attending to the technical content and accuracy but also to the human responsibility for and human consequences of how information is collected, analyzed, communicated, and used.

Secondly, statistics and data science education is an opportunity to promote an intellectual humility regarding the limits of one's knowledge. Statistics is sometimes called the "science of uncertainty" (e.g., Evans & Rosenthal, 2010), and statistical reasoning is fundamentally about variability and incomplete information. Stewardship of information requires more than drawing conclusions from data; it involves appropriately accounting for uncertainty, recognizing the appropriate scope of evidential claims, communicating limitations, and remaining open to revision in light of new information. Whether interpreting a confidence interval, evaluating a model, making data-informed decisions, or communicating findings to stakeholders, responsible stewardship of information involves resisting the temptation to overstate what is known. Statistics education at its best, then, has the opportunity to form not only professionals who can deal appropriately with statistical uncertainty but who also demonstrate an intellectual humility and sophisticated posture towards evidence and uncertainty in their personal and civic lives as well.

Taken together, we believe statistics and data science education are fertile ground for student formation that extends far beyond content mastery. The statistics and data science education literature have long emphasized student learning goals involving data & statistical literacy, evidence-based decision-making, communication, civic responsibility, and responsible and ethical use of data. However, while we may agree these goals are worthwhile, they can often feel lofty or unattainable. Practical challenges instructors may face include feeling like they do not have: 1) the bandwidth to overhaul their curriculum, 2) time to fit these goals into their courses alongside statistical content, 3) adequate training to navigate ethics or civics discussions, or 4) concrete resources for where to start. In the remainder of this paper, we aim to demonstrate how articulating *stewardship of information* to students as an overarching and repeated theme can provide a powerful and flexible framework for advancing many of these goals throughout our curriculum.

## 3. Overarching Pedagogical Approach & Implementation Strategy

To integrate stewardship of information as a formational pursuit throughout a statistics and data science curriculum, we seek to 1) introduce the theme early, 2) continually revisit the theme within and across most courses, 3) strategically align the theme with course goals and audience, and 4) developmentally increase the depth and sophistication of student engagement with the theme across courses. This approach has been informed by recommendations in the literature for ethics integration throughout computer science and data science curriculum as well as the pedagogical technique known as spiral learning (Carter & Crockett, 2019; GAISE College Report ASA Revision Committee, 2025; Harden, 1999; Kugler, 2025; National Academies of Sciences, Engineering, and Medicine, 2018). The pedagogical ideas and curriculum discussed here have been implemented by the author in multiple institutional contexts including a small Hispanic Serving Institution on the West Coast as well as a mid-sized private university on the East Coast. This paper will focus on implementation in undergraduate courses, though we have used a similar approach in graduate courses as well.

We begin every course we teach by introducing the themes of stewardship of information and data for social good on Day 1, and even in the syllabus. Example syllabus language is provided in the Supplementary Material, and Day 1 activities are discussed in Section 4. While conceptually we consider "data for social good" to be one manifestation of the overarching "stewardship of information," we believe the former to be a pedagogically powerful phrase in its own right, so we often articulate these to students as two complementary themes.

Across all our courses, our primary pedagogical workhorse for having students engage with these themes is via weekly readings, videos, or podcasts[1] outside of class. While instructors may have varied approaches for having students engage with readings (e.g., discussion posts, entrance tickets), we have found high pedagogical value in the tool Perusall, which is a free online "social learning platform" or "collaborative annotation tool" that allows students to highlight and add comments or questions to a common document, video, webpage, or podcast uploaded by the instructor. Having used Perusall for several years across a variety of course types and student populations, anecdotally we have found several pedagogical advantages. First, students engage more regularly and substantively with each other's ideas than they do in discussion posts, because they can see each other's comments in the document as they are reading. Second, the annotation feature encourages students to respond to specific passages they read rather than compose a polished synthesis at the end. As an instructor, these more informal in-the-moment reflections can often provide revealing insight into students' thinking and reactions in ways that polished discussion posts may obscure. In this same way, students seem less prone to resort to AI-generated responses, because formal summary paragraphs are not expected, and the annotation feature structurally signals to them that their genuine free-form thoughts are encouraged.

[1] For brevity, we will refer simply to "readings" moving forward, but the reader should note this could include video or audio mediums as well.

Perusall allows rich engagement with stewardship themes to happen outside of class without "costing" coveted in-class time. In a 15-week semester, we typically assign Perusall annotations in 10 of the weeks, and we expect each assignment to take students between 30 minutes and 1 hour. We supplement this with in-class discussion or activity a few times per semester, depending on the course. Because engagement is happening weekly outside of class, students have a breadth of exposure and percolating thoughts and questions that enable in-class time to move more quickly towards richer discussion.

We have found the general infrastructure of weekly Perusall readings to be a low-barrier, flexible approach that creates significant runway to meaningfully engage students with the sorts of ideas we believe contribute to their ethical, civic, and professional formation. In the next section, we discuss specific readings and activities used in each course type and how they contribute to such formation.

## 4. Implementation Across the Curriculum

### 4.1 Day 1 in Each Course

We start each semester with a short set of personal slides with photos that showcase answers to four questions as an instructor: (1) who are my people, (2) where do I call home, (3) what are some things I love, and (4) what are some things I'm passionate about. For the last question, we always include, among other things, "Data for social good" and "Becoming good stewards of information." This allows us to briefly introduce these to students as a meaningful way of understanding statistics and data science and to articulate the type of formation we hope occurs during our time together that semester. This signals that students can expect the course to be more than about learning statistical content or gaining technical skills. Especially for non-majors who are taking statistics as a required course, this can serve as a hook on Day 1 and plant the seed that data might be more relevant to their lives and the things they care about in the world than they first anticipated. For majors, it invites them to consider how their career might connect to more than just their academic skills and interests. After introducing ourselves in this way, we then use the same set of questions to have students get to know each other (see Figure 2).

Beyond introducing the theme of stewardship of information, pedagogically this is intended to humanize the learning environment by humanizing ourselves as instructor, facilitating peer interactions, and signaling that we recognize students are whole people and that all aspects of their identity are welcome in the classroom. By letting students see our own passion for statistics and data science and the good it can do in the world, we signal that we care about their holistic development and that their passions and concerns in the world have a place in the classroom, too.

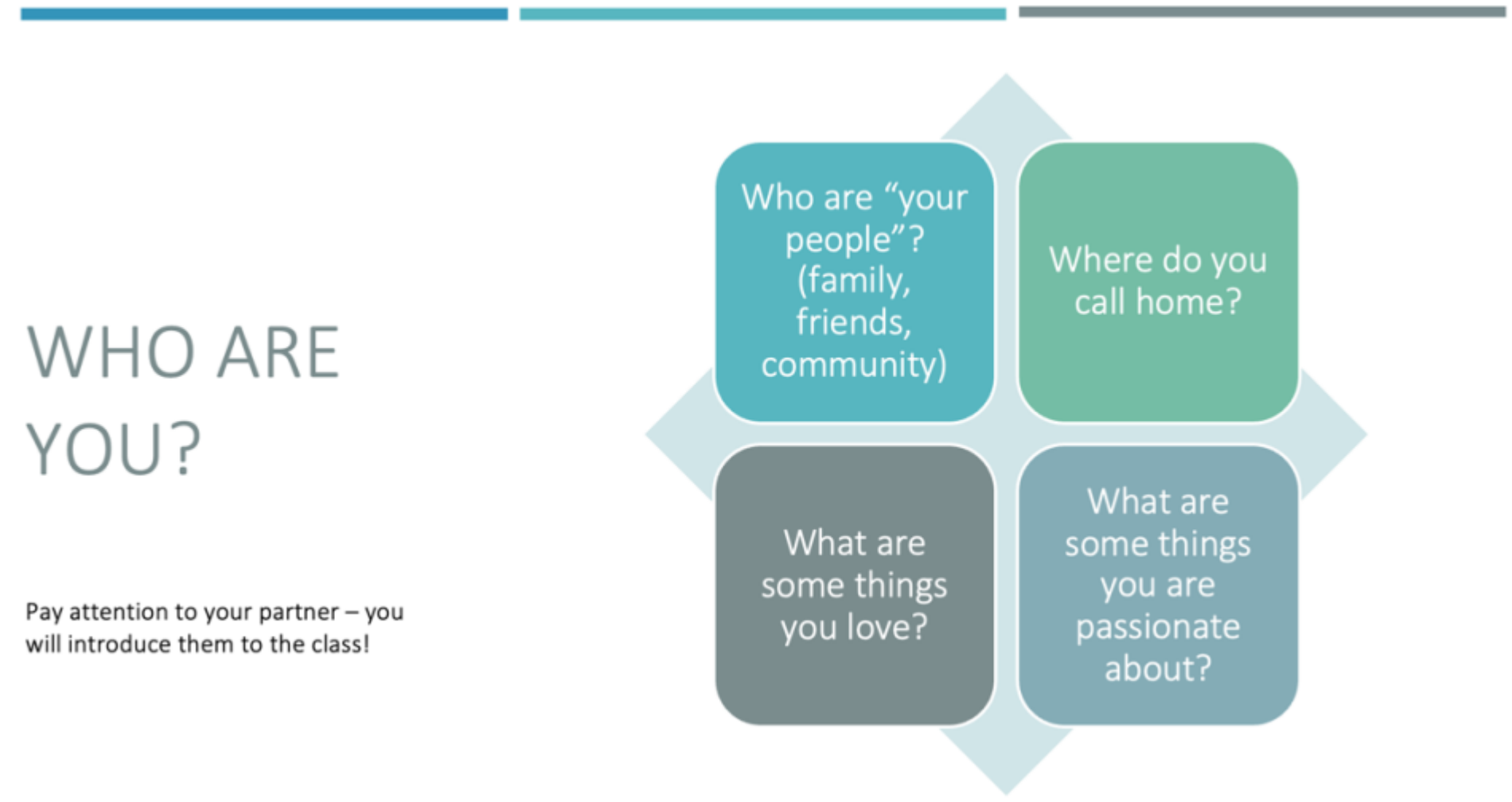

*Figure 2: Day 1 ice-breaker prompts*

**4.2 Introduction to Statistics**

In our context, Introduction to Statistics is a general education course for majors including Kinesiology, Psychology, Social Work, Theater Arts, and many more. Often, these are students who have very little if any interest in statistics or data science coming into the course and tend to exhibit high levels of anxiety about math, statistics, and/or coding. Perusall readings serve as a primary vehicle for engaging students with themes of stewardship of information and data for social good. Specifically, in this context, these readings are intended to:

1) Help students discover the relevance of data to their lives
2) Promote data literacy, especially through visualization and popular media
3) Equip students to think critically about how data is (mis)used in society
4) Use data as a tool to deepen students' concern for their community and marginalized populations

A full list of 10 Perusall assignments used in Introduction to Statistics can be found in Table 1, along with the dimensions of stewardship from Figure 1 they engage with. Collectively these readings demonstrate to students repeatedly how data can be stewarded for good; that is, that data has the power to illuminate problems and point towards solutions both at an individual level and in the public and private sectors. Examples include collecting and broadcasting early-warning data for natural disasters to reduce death tolls; holding governments accountable for deforestation; reporting location of sexual harassment incidents; monitoring the spread of disease; identifying successful poverty reduction efforts and where to allocate resources; and illuminating racial inequities in maternal child health, policing, and incarceration.

*Table 1: Perusall assignments for Introduction to Statistics*

| Week | Title | Outlet/medium (citation) | Primary stewardship dimension(s) |
|---|---|---|---|
| 1 | Data for Better Lives | World Bank blog post + video (Cull et al., 2021) | Data for social good, ethical use, human dimension |
| | To close the gender gap, we have to close the data gap | Medium blog post + video (Gates, 2016) | |
| 2 | Will you graduate? Ask big data | NYT article (Treaster, 2017) | Data for social good, ethical use, human dimension |
| 3 | The Beauty of Data Visualization | TED talk (video) (McCandless, 2010) | Communication, data for social good, literacy |
| 4 | Own Your Body's Data | TED talk (video) (Williams, 2014) | Communication, literacy, human dimension |
| 5 | Statistics, civil rights, and the U.S. Supreme court: A cautionary tale | Significance article (Kiefer et al., 2016) | Literacy, ethical use, human dimension, data for social good |
| 6 | Vast new study shows a key to reducing poverty: More friendships between rich and poor | NYT article, interactive visualizations (Miller et al., 2022) | Data for social good, human dimension |
| 7 | What do vaccine efficacy numbers actually mean? | NYT article (Zimmer & Collins, 2021) | Communication, literacy, data for social good |
| | Who had COVID-19 Vaccine Breakthrough Cases? | NYT article (Aufrichtig & Walker, 2021) | |
| 8 | Birth of a Word | TED talk (video) (Roy, 2011) | Data for social good, ethical use, human dimension |
| 9 | Data for the Public Good | NYT Op-ed (Shihipar, 2019) | Data for social good |
| 10 | 3 Ways to Spot a Bad Statistic | TED talk (video) (Chalabi, 2017) | Literacy, human dimension |

Once students see the power of data, they are more invested in the idea that it must be stewarded well. Many of the readings are also intentionally selected to help students see the relevance of data to their own lives, encouraging them to reflect on various dimensions including how they can utilize their own data (e.g., Williams, 2014), how to draw insight from data about their communities or experiences (e.g., Miller et al., 2022), and how data about them is being collected, used, and shared (e.g., Treaster, 2017). This curated set of readings also has students grapple with the consequences of data *not* being stewarded well. For example, Melinda Gates discusses how lack of data on women and girls in developing countries hampers efforts to achieve gender equity (Gates, 2016). Several articles raise a common ethical tension regarding privacy and surveillance (Cull et al., 2021; Roy, 2011; Treaster, 2017); data can be powerful, but at what point does it become invasive? For example, should citizenship data be collected on the Census? Should a university track student ID swipes? Should a researcher collect video data of his child's development? In all cases, who has access to and ownership of the data? For what purposes will the data be used? When is consent required?

Other readings have students grapple with data and statistics as evidence (Chalabi, 2017; Kiefer et al., 2016; Treaster, 2017; Williams, 2014; Zimmer & Collins, 2021). Who gets to decide what data and evidence is relevant? What happens when evidence is ignored? How much is uncertainty or variability taken into account? Interestingly, several articles unexpectedly led students to wrestle with the tension between aggregate patterns, individual experiences, and human agency. For example, when reading about the power of big data to predict student drop-out (Treaster, 2017), students often muse "Are my outcomes predictable? Am I just a statistic?" Similar tensions emerge when students grapple with aggregate evidence about vaccine efficacy and their personal anecdotes regarding vaccines and COVID (Zimmer & Collins, 2021). Talithia Williams' TED talk leads to productive discussion of this tension; Williams' medical doctor bases health recommendations on aggregate evidence, but she has personalized data she feels points towards a different conclusion and uses it to advocate for different medical care. While a similar aggregate versus individual tension is present, interestingly, Williams does not have to rely on anecdote but instead has her own finer-grained data to empower her decision-making (Williams, 2014). Notably, female students of color consistently comment how they find this example particularly empowering, and it also tends to resonate with athletes or those studying kinesiology, athletic training, or medicine.

Collectively, by engaging with these ideas and questions each week, students begin to understand that statistics is more than simply mathematical or technical expertise but rather an opportunity to steward data in service of people and communities. About two-thirds of the way through the course, we have an in-class discussion to tie together stewardship themes that students have seen throughout the course. They are then asked to write a 500-word reflection synthesizing how their understanding of data, statistics, and stewardship of information may have evolved over the course of the semester.

**4.3 Probability**

In our contexts, Probability is taken by STEM majors including mathematics, engineering, and computer science, and requires Calculus 2 or 3 as a pre-requisite. Many of these students have not taken a prior statistics course, and there is almost no overlap with students who take Introduction to Statistics. As a result, the course serves as an introduction to stewardship of information for a different audience and through a different mathematical lens. We therefore seek to:

1) Expose students to real-world applications of probability and statistics that promote social good and/or evidence-based decision-making
2) Emphasize the human dimension of often abstract mathematical content
3) Equip students to grapple with uncertainty and randomness in meaningful ways
4) Introduce students to statistical epistemologies

These goals shape our reading selection. Compared to Introduction to Statistics, we more frequently utilize articles from *Significance* magazine which tends to have a higher yet still accessible level of mathematical sophistication. Because probability courses focus less on data analysis and more on mathematical foundations, readings centered solely on data for social good are less prominent. Because course content can often feel theoretical and disconnected from real-world application, we prioritize articles that can illuminate the practical significance of abstract probabilistic ideas, and we attempt to connect readings to weekly course content when feasible (see column 4 in Table 2).

*Table 2: Perusall assignments for Probability*

| Week | Title | Outlet/medium (citation) | Statistical content | Stewardship theme(s) |
|---|---|---|---|---|
| 1 | Statistics in court: Incorrect probabilities | Significance magazine article (Divine, 2019) | Probability, independence, multiplication rule | Literacy, ethical use, communication |
| 2 | Desperately seeking signal, chapter in The Signal and The Noise | Book chapter (Silver, 2015) | Randomness, Models, Uncertainty | Literacy, uncertainty |
| | Chance, chapter in Mathematics Through the Eyes of Faith[2] | Book chapter (Howell & Bradley, 2011) | | Uncertainty, human dimension |
| 3 | Bayes Theorem: The geometry of changing beliefs | YouTube video (3Blue1Brown, 2019) | Bayes Theorem | Literacy, uncertainty, human dimension |
| | From Evidence to Causes: Reverend Bayes Meets Mr. Holmes, chapter from The Book of Why | Book chapter (Pearl & Mackenzie, 2019) | | |
| 4 | The shock of the mean | Significance magazine article (Raper, 2017) | Expected value, aggregation | Literacy, human dimension, communication |
| 5 | William Guy: Victorian statistics' biggest champion | Significance magazine article (Bennett, 2025) | Poisson distribution | Data for social good, communication |
| | Are heatwaves more deadly for women? | Significance magazine article (Mukherjee, 2024) | | Human dimension, uncertainty, data for social good |
| 6 | Preventing cancer: mere rhetoric or a promising plan? | Significance magazine article (Valberg, 2019) | Probability, expected value, variance, Lorenz curve | Uncertainty, data for social good |
| 7 | Weapons of Math Destruction: Intro and Chapter 1: What is a model? | Book chapter (O'Neil, 2016b) | Models | Ethical use, human dimension, data for social good |
| 8 | SAT to Give Students 'Adversity Score' to Capture Social and Economic Background | Wall Street Journal article (Belkin, 2019a) | Normal distribution | Ethical use, human dimension, communication |
| | College Board drops plans for SAT Adversity Scores | Wall Street Journal article (Belkin, 2019b) | | |
| 9 | Vast new study shows a key to reducing poverty: More friendships between rich and poor | NYT article, interactive visualizations (Miller et al., 2022) | Correlation, regression | Data for social good, human dimension |
| 10 | Will you graduate? Ask big data | NYT article (Treaster, 2017) | Probability, prediction, uncertainty | Data for social good, ethical use, human dimension |

As with Introduction to Statistics, the readings are chosen to expose students to a breadth of applications. Here, application areas include criminal justice, earthquake forecasting, public health, medicine, climate change, genetics, and education. Students are often surprised by the case studies in "Statistics in Court: Incorrect Probabilities" that highlight how courts have made decisions based on incorrect understanding of the very basic probability principles students have learned in week 1 of the course, such as independence and the multiplication rule. This poor statistical literacy has led to wrongful convictions of innocent defendants (Divine, 2019). Starting with this reading emphasizes the importance of statistical literacy early on and sets a humanizing tone for what can otherwise be dry or abstract set of mathematical course objectives.

[2] This reading is used in a faith-based institution but can be made optional or excluded in a secular context. It has students wrestle with philosophical dimensions of randomness, such as the ontological existence of chance and its implications for the nature of God and free will.

Other readings also focus on human implications, such as the extent to which a students' social and economic adversity should play in college admissions (Belkin, 2019a, 2019b), the disproportionate toll of heatwaves on women (Mukherjee, 2024), individuals' risk of getting cancer (Valberg, 2019), and how models are used (ethically or unethically) to determine life-altering outcomes for individuals, such as whether they should get hired, fired, approved for a loan, or sent to prison (O'Neil, 2016b). Two of the more student-relevant articles from intro stats are included here as well: "Will you graduate? Ask big data" and one about friendships between rich and poor as a key to reducing poverty, which tends to engender reflections on their own friendships and where they see themselves in the data (Miller et al., 2022; Treaster, 2017).

Many of the articles discuss statistical uncertainty and concepts of randomness and variability, but the readings in Weeks 2 and 3 have students engage with these ideas more deeply and philosophically. Chapter 5 of Nate Silvers' *The Signal and The Noise* provides a compelling discussion of the practical difficulties scientists face in making sense of uncertainty in the context of earthquake predictions and how an underappreciation of uncertainty and poor probabilistic reasoning can lead to disastrous real-world consequences (Silver, 2015). In institutional contexts where faith-related discourse is welcome, the chapter *Chance* in the book *Mathematics Through The Eyes of Faith* generates thoughtful discussion, as students grapple with randomness on both a human and ontological level and whether they view that as compatible or incompatible with their notions of a sovereign deity (Howell & Bradley, 2011).

Finally, we believe these readings help introduce students to statistical epistemologies; that is, the habits of mind and established ways of knowing that have been adopted by our profession. Having explicit discussion about these habits of mind contributes to their professional formation. Critical evaluation of randomness and uncertainty, Bayesian belief-updating, and the power of aggregation are examples of key statistical epistemologies that each find relevance in a probability course. A 3blue1brown video on Bayes Theorem as "the geometry of changing beliefs" and a book chapter from *The Book of Why* introduce students to the broader world of Bayesian statistics and its applications as well as to the epistemological underpinnings of uncertainty as degree of belief and Bayes Theorem as a framework for updating beliefs in principled ways in the face of new evidence (3Blue1Brown, 2019; Pearl & Mackenzie, 2019). Relatedly, the Significance article *The Shock of the Mean,* details the history of the statistical average and how it was surprisingly controversial to suggest that aggregating observations could prove informative. As the article points out, the mean is so commonplace now that its usefulness seems obvious and its existence inevitable. This reading and others (on Bayes and William Guy) push students to consider that these methods have not always existed; rather, humans drive methodological development and come to (sometimes implicit) consensus about epistemological norms (Bennett, 2025; Pearl & Mackenzie, 2019; Raper, 2017). We hope this plants seeds that an academic discipline such as mathematics or statistics is not just an abstract set of axioms and formulas that exist in the world but is a dynamic community of humans who push thinking forward. For students who continue in their statistics education, we push further on this idea of statistical epistemologies and their human dimension in upper division courses, to be discussed in Section 4.5.

**4.4 Data Science**

Our Data Science course is primarily taken by students pursuing a major concentration or minor in statistics or data science, and at least one prior statistics course is required. This means that unlike a general education course or a required STEM probability course, these students have a demonstrated interest in data science and may even pursue it as a career pathway. Consequently, our stewardship goals shift from introducing students to the value of data towards forming them as thoughtful and ethical data practitioners. Through a series of Perusall readings in this course, we seek to:

1) Engender interest in data science as a meaningful career
2) Contribute to students' formation as ethical data scientists
3) Facilitate student reflection on (de-)humanization through quantification
4) Challenge students' notions that "the numbers don't lie" or "the data speaks for itself"

The chosen sequence of readings heavily emphasizes the responsibility students hold as budding data scientists and the ethical tensions they should be mindful of in the collection, analysis, communication, and use of data (see Table 3).

These readings still point towards how data can be used for social good, but now much more attention is paid to examples of *poor* stewardship of information. Stated differently, these readings push students to recognize just how much intentionality and pro-active care is required to steward data for good. Students begin to wrestle with questions such as what should happen to data once it has been used for its intended purpose? Does the value of the data to accomplish one goal outweigh potential risks if used for other goals? For example, Spotify can argue that access to a person's photos, contact lists, microphones, and locations can help them give better personalized music recommendations, but does a company having that amount of information pose other risks? (Onuoha, 2016) Or what happens when sensitive data that has a legitimate use by one sector of the government is shared with another branch of government? Students read about the Census Bureau's later admission that, during World War II, it shared names and addresses of Japanese Americans with the Secret Service; this data was subsequently used to help facilitate the roundup and imprisonment of these U.S. citizens in internment camps (Minkel, 2007). Students also read of more modern examples of the 2025 U.S. Department of Government Efficiency (DOGE) consolidation of sensitive data from many government sources and in some cases sharing it with Immigration & Customs Enforcement (ICE) (Blake, 2025; Chishti & Putzel-Kavanaugh, 2025; Kindy & Seitz, 2025). Students grapple with who owns data, when it should be shared, and tensions between privacy, surveillance, and community safety or national security.

*Table 3: Perusall assignments for Data Science*

| Week | Title | Outlet/medium (citation) | Primary stewardship theme(s) |
|---|---|---|---|
| 1 | Visualizing Progress: Data Insights from the 2023 ATLAS of Sustainable Development Goals | YouTube video (first 15 minutes) (World Bank, 2023) | Communication, data for social good |
| | Data for the Public Good | NYT Op-ed (Shihipar, 2019) | |
| | The War on Data | Chance magazine article (Gelman & Palko, 2013) | |
| 2 | The Point of Collection | Medium blogpost (Onuoha, 2016) | Human dimension, literacy |
| | On Missing Data Sets | GitHub webpage (Onuoha, 2017) | |
| | “Raw Data” is an Oxymoron | Book intro (Gitelman & Jackson, 2013) | |
| 3 | The Ethical Data Scientist | Slate article (O’Neil, 2016c) | Ethical use, human dimension, data for social good |
| | Skulls and Skin (Seeing White, Part 8) | Scene on Radio podcast (Biewen & Kumanyika, 2017) | |
| | Beyond Racial Statistics & Challenging Race as a Variable, chapters from Thicker Than Blood | Book chapters (Zuberi, 2001) | |
| 4 | Weapons of Math Destruction Talks at Google | YouTube video (O’Neil, 2016a) | Ethical use, human dimension |
| 5 | To close the gender gap, we have to close the data gap | Blogpost (Gates, 2016) | Human dimension, data social good |
| | Can snow-clearing be sexist? Chapter in Invisible Women: Data Bias in a World Designed for Men | Book chapter (Perez, 2019) | |
| 6 | Mistakes, we’ve drawn a few: Learning from our errors in data visualization | The Economist article (Leo, 2019) | Communication |
| | Upping Your Data Viz Game | YouTube video (Clause Wilke & Santa Fe Institute, 2020) | |
| 7 | Cambridge Analytica Whistleblower | YouTube video (The Guardian, 2018) | Ethical use, human dimension |
| | The Ethics of Using Hacked Data: Patreon’s Data hack and academic data standards | Data & Society case study (Poor & Davidson, 2016) | |
| 8 | DOGE’s access to Education Department Data Raises Concerns | Inside Higher Ed article (Blake, 2025) | Ethical use, human dimension, data for social good |
| | Confirmed: The U.S. Census Bureau Gave Up Names of Japanese-Americans in WWII | Scientific American article (Minkel, 2007) | |
| | Seeking to Ramp Up Deportations, the Trump Administration Quietly Expands a Vast Web of Data | Migration Policy Institute article (Chishti & Putzel-Kavanaugh, 2025) | |
| | Trump administration hands over nation’s Medicaid enrollee data to ICE | Associated Press article (Kindy & Seitz, 2025) | |
| 9 | The Data Divide is Real, and could be highly destabilizing | Forbes article (Splunk, 2022) | Data for social good, ethical use, human dimension |
| | Closing the Data Divide for a More Equitable U.S. Digital Economy | Center for Data Innovation report (Diebold, 2022) | |
| | Inside the race to archive the government’s websites | MIT Technology Review article (Mulligan, 2025) | |
| 10 | The Nation’s Data is at Risk (Executive Summary) | American Statistical Association report (American Statistical Association, 2024) | Data for social good |
| | The Threat to Federal School Data is a Threat to Us All | EducationWeek Op-ed (Wasserstein, 2025) | |

We assign a pair of readings by Mimi Onuoha, a visual artist and writer whose work examines the effect of data and technology on society, who demonstrates that datasets, especially when combined, can reveal far more than intended (Onuoha, 2016). She quotes Bruce Schneier in his book *Data and Goliath* as saying "data we're willing to share can imply conclusions that we don't want to share" (Schneier, 2016). Onuoha's work also elegantly pushes students to recognize the humanity that is infused in every dimension of data (Onuoha, 2016, 2017). Her writings, paired with "Raw data is an oxymoron" (Gitelman & Jackson, 2013) surface the reality that not only is data often collected *about humans* (sometimes, at great cost, as in the case of sexual assault incidents), but perhaps more importantly, data is collected *by humans*. In class, we emphasize that the research questions we ask, the data we collect, and the way we use that data are infused with (often hidden) values about who and what matters in the world. As Onuoha puts it, "To collect, record, and archive aspects of the world is an intentional act, one that typically benefits those who have the power to decide what should be collected" (Onuoha, 2016). In this vein, we have students reflect on the following questions in various assignments:

1. Who benefits from this data/analysis? Who is potentially harmed?
2. Whose data/perspective is included? Whose is left out?
3. What are the potential consequences of this data/analysis (good and bad)? Think of stakeholders at various levels.

Onuoha (2017) pushes students to examine what datasets are missing and what incentive structures might be preventing their existence (Onuoha, 2017). Related readings have students reflect on gender bias in data as well as "the data divide" that often falls along socioeconomic and racial/ethnic lines (Diebold, 2022; Gates, 2016; Perez, 2019; Splunk, 2022). This all lays important foundation for students to engage more deeply at the end of the semester with the implications of the weakening federal data infrastructure (American Statistical Association, 2024; Mulligan, 2025; Wasserstein, 2025).

Many assigned articles touch on race as well as algorithmic bias, but Week 3 tackles the role of race in statistics and data science head-on. The podcast episode "Skulls and Skin" gives students historical context for the role science and data has played in classifying and quantifying race in often problematic ways and how that quantification gave a veneer of objectivity to a definition that is socially constructed (Biewen & Kumanyika, 2017). In data science, it is commonplace to encounter data sets with a race/ethnicity variable, but rarely are the implications of its existence or its use given much thought. Work by Cathy O'Neil presses students to consider instances where considering race in a model may introduce harmful bias, even as it would provide "useful" predictive power (O'Neil, 2016a, 2016c). Two book chapters, *Beyond Racial Statistics* and *Challenging Race as a Variable,* push students to wrestle with whether measuring race fundamentally reifies harmful social constructs and racial hierarchies or whether it is still a net-good because of its role in illuminating injustices and holding institutions accountable for inequity (Zuberi, 2001). While we do not resolve these tensions or come to prescribed guidelines for how to universally handle race as a variable, we hope these discussions contribute to students' formation as thoughtful data scientists who are aware enough to pause, think through, and seek input on the implications of their decisions in any given context.

The readings for this course engage with more mature and politically adjacent themes including race, gender, and government use of data. As such, it is crucial to create a respectful and inclusive learning environment conducive to these potentially sensitive conversations. Beyond our standard Day 1 activities and overall pedagogical techniques for creating a student-centered learning environment, we also co-create classroom norms on Day 1. These vary by semester, but we often nudge students to include versions of the following norms[3] pertinent to sensitive discussion: everyone has expertise; critique ideas, not people; engage new perspectives with curiosity, not judgment; exhibit intellectual humility; normalize time to think; and embrace discomfort and non-closure. Students are reminded of these norms throughout the semester.

During the 1st week, we also do a Social Identity activity proposed by Dr. Missy Crosby (2022) to reflect on and engage in discussion about how our social identities affect our positionality in a mathematics or statistics classroom. The activity involves each student writing aspects of their identity on 4 notecards: (1) gender, (2) race/ethnicity, (3) religious/spiritual affiliation, and (4) an additional aspect of their choosing that reflects something about who they are (e.g., major, hobby, familial identity, political affiliation). They are then presented with a series of places: (a) home, (b) on-campus but outside of classroom, (c) inside a mathematics or statistics classroom, and (d) other off-campus location frequented in the community (e.g., gym, grocery store, place of worship). For each location (a) – (d), students are asked to place their identity cards (1) – (4) in order of which aspect of their identity they are most aware of in that location. After each location, students discuss in small groups why they chose the ordering they did[4]. At the end of all 4 locations, there is a whole class discussion. This elicits rich conversation about our social identities in different spaces, why some identities feel more safe or exposed or threatened in certain spaces, and how our experiences may be different from each other. This activity sets the stage for ongoing reflection questions students engage with throughout the semester regarding how our social identities affect the lens through which we interpret data as well as inform the questions we ask, the data we collect, and the analyses we conduct in the first place.

Because data ethics is a more naturally relevant topic in a Data Science course, we spend a full week of class time discussing ethics topics in more depth. There are many excellent resources for data ethics modules and activities in the literature, so we only highlight a sequence of activities here to demonstrate how we utilize data ethics in service of the broader formational aims of helping students become good stewards of information.

[3] These norms were inspired by and adapted from recommendations provided by Project NExT (Kung, 2022)
[4] It should be emphasized that students do not need to discuss any aspects of their identity they do not feel comfortable disclosing.

We usually time the ethics module to come after the readings in Weeks 7 or 8 of Table 3, so that students have a breadth and depth of exposure to potential ethical issues. We begin with a brief warm-up that reminds students of the different ethical issues they have encountered so far in the course (e.g., bias & representation, missing datasets, data privacy, (mis)use of data or surveillance by corporations or government, visual misrepresentation of data, and weakening of federal data infrastructure), and they are given a few minutes to do a free-write about which ones they are most concerned about and why. After a brief paired and whole-class discussion, we then post the question “Why might different people worry about different issues?” We briefly recall the social identity activity as a reminder of how our identity shapes our worldview and experiences.

We then turn to a Core Values activity, adapted from the Integrated Ethics Labs (Carter, 2021), in which students are presented with a list of values (see Supplementary Material) and told to circle 20-25 that are most important to them. Then, in a secondary pass, they place a W next to values that might influence where they might choose to work when they graduate. The pedagogical aims of this activity are threefold: 1) it requires students to explicitly reflect on their values, which is not often done in educational settings, 2) it has them consider how their values connect to their career choices, which is important for ethical and professional formation, and 3) it illuminates that different people have different values. In particular, it is interesting for students to notice that there is near universal agreement that each of the values is good individually, yet each person is drawn to a different subset of those values. And different prioritization of values will lead to different decisions. Crucially, this illuminates how many tensions arise in data science collaboration and society at large, because all values cannot be optimized at once and in fact may often be in direct tension with one another. For example, privacy may be at odds with transparency, or freedom may be in tension with safety or security. Whether to use race in an analysis if often a tension between usefulness and fairness, or efficiency and dignity. When considering government use of data, tensions between privacy and surveillance abound, and surveillance is often justified as being in service of other values such as safety, security, or justice. We believe these conversations to be important for engendering a posture of intellectual humility, curiosity, and generosity towards those who may think differently from us.

The values activity leads directly into discussion of data science case studies such as bias in predictive policing, targeted political advertising to influence groups of voters, social media experiments to influence mood or behavior, and scraping social media images to build facial recognition software used for surveillance. We find that students are often able to easily identify bias and how things have gone awry in retrospect, but sometimes they land on simplistic explanations that assume only nefarious motives and actors are to blame. While this is sometimes an accurate diagnosis, we push students to consider what perhaps well-intentioned goals may have existed at the start of a project. For example, imagine you are a data scientist hired to by a political campaign and your goal is to get your candidate elected. Or, you are hired by law enforcement with the goal of using data to preventing crime and make neighborhoods safer. Or, more generally, you are hired to use data to increase your company's profit or market share. Our intention is not to normalize or excuse harmful consequences of data use but rather to help students realize that most systems begin with at least a loosely justifiable or even positive goal. Further, as future data scientists, they, too, are capable of creating unintentionally harmful data systems. In other words, the goal is to emphasize that ethical data science does not happen by default; it is done intentionally by people who take responsibility for building better systems.

**4.5 Upper Division Courses for Statistics Majors or Minors**

Here, we discuss implementation of stewardship themes in upper division courses for statistics majors or minors such as Mathematical Statistics or Statistical Models. Importantly, the students in these courses have already taken at least two statistics courses – often the Probability and Data Science courses above – and are in a degree program that requires not just data science but more advanced statistical methods. As such, we view this as an appropriate time in their statistical education to invite them into an "in-house" conversation – as fellow statisticians – about the complicated history of statistics as intertwined with racism and eugenics. These students are typically juniors and seniors, so these complex conversations are more developmentally appropriate at this stage, as well. Our formational goals for students in this course are to:

1) Enhance their skill of asking good (statistical) questions
2) Deepen reflections on (de)-humanization through quantification
3) Recognize, employ, and critique statistical epistemologies

We believe that question-asking is an important formational aspect of becoming a good statistician. Much of statistics is asking well-formulated questions that can be answered with data. We also want students to begin questioning the "basic" assumptions they may have previously accepted as fact. For example, when it comes to statistical procedures, norms, and interpretations, it is prudent to ask: *Why is this true? Under what circumstances is this true? What would happen if…?*

To this end, we begin Day 1 with a question-generating activity inspired by a Project NExT session held by Stephanie Salamone (2022), where the instructor projects a quote on the screen (ideally one that can provoke rich food-for-thought relevant to course content), and students are given sticky notes and asked to generate 15 – 20 questions about the quote in a 5 – 7 minute time span. This is an intentionally hard number of questions to achieve; participants become painfully aware of the question-asking process itself and begin to ask deeper questions as well as more trivial ones than they otherwise would have. In general, this activity – agnostic to the quote itself - is intended to set the tone at the start of the course that frequent question asking is not only encouraged but expected in the course. In this context, we choose the following quote in service of our formational goals for this course:

> *"Statistics, as a lens through which scientists investigate real-world questions, has always been smudged by the fingerprints of the people holding the lens."*

The quote comes from an article by Aubrey Clayton, *How Eugenics Shaped Statistics,* that students read in full later on in the semester and is intended to set the stage for the semester-long conversation about how statistics has been tainted by its eugenicist roots and what that means for us as modern statisticians (Clayton, 2020). This and other key assigned readings are provided in Table 4. Note that in our upper division courses, students are provided with a large repository of journal articles from the applied statistics for human rights literature, and they each choose the assigned readings for the class in the latter half of the semester. Thus, the number of instructor-assigned readings is fewer.

*Table 4: Perusall readings in upper division courses*

| Title | Outlet/Medium (citation) | Stewardship theme(s) |
|---|---|---|
| How Eugenics Shaped Statistics | Nautilus article (Clayton, 2020) | Human dimension, ethical use |
| Why was Galton so concerned about "regression to the mean"? A contribution to interpreting and changing science and society | Journal Article, DataCrítica: International Journal of Critical Statistics (Taylor, 2008) | Human dimension, literacy |
| Statistical Methods for Human Rights, selected chapters | Book chapters (Asher et al., 2008) | Human dimension, data for social good, literacy, communication |

Through these readings, we wrestle with the fact that not only were many of the founding fathers of statistics eugenicists (including Francis Galton, Ronald Fisher, and Karl Pearson), but in some instances their statistical methodology was developed precisely in service of trying to prove racial difference and hierarchy. These statisticians and their methods played an important role a broader 20th century movement that attempted to legitimize eugenics as an objective or scientific field. Clayton (2020) argues that while we may be able to remove names from buildings, awards, and monuments, there remain eugenicists roots in the language, logic, and philosophy of statistics itself. This invites further conversation about statistical epistemology that was introduced in Probability. What are the ways of thinking, knowing, and discerning truth that have been established by (people in) our field? When are those valid and when are they deserving of critique?

Excerpts from the book *Statistical Methods for Human Rights* provide more positive – yet still morally and philosophically complicated – examples of statistical epistemologies in practice (Asher et al., 2008). In much the same way that any applied statistics journal article might, these readings expose students to real-world examples of how researchers go about turning data into evidence, how they craft their arguments, what assumptions are made, and whether those arguments and assumptions are believable. Through seeing several applied journal articles, students begin to identify what types of methodological choices and reporting strategies come across as more or less credible. As such, they are discerning what it means to be a good steward of information in both analysis and communication.

Statistical work in the field of human rights work is often conducted under significant practical constraints that create both methodological and ethical challenges for statisticians. Through real-world human rights case studies, students are exposed to advanced statistical methods including sampling, missing data, record linkage, statistical estimation, and modeling, and they examine challenges related to data collection of sensitive information, data quality, bias, government data infrastructure, privacy, and transparency. Human rights applications include topics such as missing females, deaths of combatants and non-combatants in war, criminal justice, violence against women, human trafficking, profiling, refugees & immigration, poverty, hate crimes, and discrimination. Students come to realize that often what ends up coded and categorized in a dataset started as written or even verbal documentation of an encounter between two or more humans.  The process of turning that information into quantifiable metrics is far from trivial and involves many decision points and assumptions. The following quotes from students in an upper division course illustrate the types of epistemological, ethical, and statistical tensions students wrestle with as they engage with the readings:

> "Without a clear definition of what constitutes trafficking and what does not, then there cannot be any reliable estimate of people who are victimized by it."
>
> "The book chapter [talks] about several different and complicated social/historical factors, gaps in data, and other difficulties that make it hard to run a statistical analysis to determine whether or not the death penalty results in a deterrence effect in homicide rates…. Because of these and potentially other limitations of the available (and maybe nonavailable?) evidence, I think it is fair to say that more than statistics needs to be involved here."
>
> "Is there even a reliable way of converting qualitative data into quantitative data?… Is there a way of taking an anecdote on instances of human rights violation and quantifying it, like they suggest in this section (pg. 18)? Lots of decisions need to be made, which I imagine might also lead to a lot of nuanced ethical concerns: what kind of information are we quantifying? What information will we ignore? Who gets to make these decisions and on what grounds? Do we not have to assume that the anecdotes are impartial? Granted these factors, how objective can our quantified data even be? Is there a possible risk of de-humanizing human rights?"

In class, we dive into deeper discussion of the readings, often returning to guiding questions:

1) What are the statistical epistemologies employed here – how do the authors go about using data as evidence? Is it believable?
2) What obligations do we have to be good stewards of information when data is incomplete or messy? How do we go about this?
3) What is the role of quantification? When can it be humanizing and when can it be de-humanizing?

Later in the course, the same student from the last quote demonstrates this progression in their thinking:

> "This shows that statistics can be used as a powerful tool to piece together pieces of the story that may not be clearly related at first. I know that I have expressed concern in some of my previous posts about the idea of quantifying the human story, but I think this chapter makes it a little clearer to me that it is very difficult in the meaningful practice of statistics to completely discard the human story once the numerical information has been gathered from it. Because if statistics are able to tell us a lot, they [do] not really say much without the **context** offered by the human story--and (with this chapter at least) they are meant to fill in the gaps not explicitly told by the human story."

This quote illustrates how the human rights context invites deeper reflection on (de)-humanization through quantification. While the eugenics conversations demonstrate that statistics have clearly been used for nefarious racist ends, the statistical human rights literature provides compelling success stories of how statistical methods have clearly been used in service of human rights as well. At the same time, reducing human rights abuses to a data point or statistic can also feel de-humanizing. We have found the readings and discussions in upper division courses to be especially fruitful in helping students reason in a thoughtful and sophisticated manner about statistical methodology and its human dimensions.

## 5. Discussion

### 5.1 Evidence of Student Engagement

While a full qualitative analysis of student artifacts is outside the scope of this paper, here we provide some additional illustrative student quotes from end-of-course reflections as preliminary qualitative insight into how students are engaging with the framework. To do so, we organize quotes in Table 5 according to the five characteristics we used to describe good stewards of information at the start of the paper. Course artifacts were collected under an exempt Institutional Review Board (IRB) protocol, and additional quotes are used with students' explicit permission. Overall, students, especially in introductory courses, report being surprised by the relevance of data to their lives and by the power data holds to accomplish good in the world. Students often identify personal takeaways for how they can be good stewards of information in their own lives and careers. Others indicate that the course raised their awareness of social issues, and that data and statistics have the power to change perceptions. Students also commonly report that they walk away from the course recognizing that statistics is more complex, interdisciplinary, and human than they first expected. Importantly, they leave recognizing statistics and data science is more than just technical competency, and more mindful of the additional skills required to be a thoughtful statistician or data scientist including communication and ethics.

Taken together, the reflections in Table 5 suggest that students are engaging with statistics and data science in ways that extend beyond content mastery. Across courses, students frequently report becoming more thoughtful consumers and communicators of information and evidence, more aware of the ethical and human dimensions of statistics and data science, and more inspired by the ways data and statistics can be used to serve communities and advance the common good. While these quotes are illustrative rather than the product of formal qualitative analysis, they suggest that stewardship of information can provide students with a framework and language for connecting statistics and data science to broader ethical, professional, and civic responsibilities and opportunities.

*Table 5: Student quotes from end-of-course reflections*

| Goal for students to become: | Illustrative quotes | Course |
|---|---|---|
| **People who are thoughtful producers and consumers of data and evidence** | “Before taking this course, I thought data literacy had to do with numbers, coding, or something in the mathematical sphere....it is so much more than that—it’s about championing the truth, learning to discern, and ultimately becoming responsible stewards of information” | Introduction to Statistics |
| | “I’ve become more conscious of how I use and share information on social media and in my academic work.” | Probability |
| | "One take-away is that I feel like I can read news articles that cite to statistics with a more critical eye and have a better understanding of what the cited results actually mean." | Probability |
| **People who exhibit integrity and transparency in communication of findings** | "There is so much non-mathematical work involved in statistics ... making sure to record and interpret data with integrity." | Probability |
| | “Being able to obtain the data is only half the story—if you cannot visualize it and communicate it to the audience in an effective ... manner, the data means nothing” | Data Science |
| | “I think I was surprised about how much communication was a part of this class, and it was exciting to learn not only how to code, but how to communicate results” | Data Science |
| **People who think critically about data ethics** | “The topics of the readings were important to consider when conducting data collection, data use, and overall data ethics. In other classes, I have not learned/discussed data ethics on this deep of a level, so there were things I never even really considered or thought about until the readings, which really opened my eyes to issues surrounding data science” | Data Science |
| | "Statistics is found at this interesting intersection between theoretical math, real-world applications, and ethical philosophy." | Probability |
| | “I believe that data ethics is one of, if not the most, important aspect when working with data. It is crucial that in the process of obtaining data, nobody is exploited or harmed, and that information is never used to manipulate others” | Data Science |
| **People who recognize the human beings behind the data** | “Managing knowledge with honesty, consideration, and purpose is what it means to be a good steward of information. I've learned in this class that data is about people, not just numbers, and that how we utilize it may either advance justice or exacerbate existing issues.” | Introduction to Statistics |
| | “It was never lost on me during this course that at the end of the day, data is not just numbers, but consequences, explanations, and solutions that affect real people” | Data Science |
| **People who harness data to make informed decisions, serve communities, and advance the common good** | “As a future healthcare provider, I will live out my stewardship by protecting patient data, sharing information that benefits patient health, and using relevant information to ensure my future patients don't become another statistic... I am better equipped to responsibly and ethically use data for the greater good of society. This course has shown me that data is more than just numbers—it can be an impactful resource for the pursuit of justice.” | Introduction to Statistics |
| | “I firmly believe that the use of statistics to understand the world has important implications, because understanding the world and why things happen can change our views towards empathy and seeking solutions” | Introduction to Statistics |

**5.2 Reflection on Implementation & Ongoing Challenges**

Overall, we have found that students engage consistently and meaningfully with Perusall assignments, and we anecdotally consider there to be significant payoff for relatively low pedagogical effort. Having weekly readings built into all our courses has provided the infrastructure to more readily implement culturally responsive pedagogy as well as facilitate engagement with current events or other timely topics. Readings are chosen to foster students' active and direct engagement with ideas from a diversity of authors and perspectives far beyond what a single instructor can offer through their own lived experience. The readings themselves are curated to showcase statisticians, data scientists, and other thinkers from a variety of gender and racial/ethnic identities. The pedagogical design of Perusall is also intended to facilitate meaningful engagement between peers, bringing an even greater diversity of perspectives into the learning environment. As seen in the final quote of Table 4, this overall approach has the potential to increase students' empathy and their understanding of the world around them.

However, there are several challenges and areas for improvement. Chiefly, knowing how to meaningfully assess these "softer" learning goals is a challenge. Typically, we have allotted 5 – 10% of the course grade to Perusall annotations and related reflection assignments. Perusall has functionality for auto-grading according to flexible criteria set by the instructor as well as integration with common Learning Management Systems. This allows for Perusall assignments to be low-friction formative assessments that ideally contribute to student learning but require minimal effort from the instructor beyond initial set-up. However, practically speaking, Perusall assignments end up being graded primarily on completion; that is, if students spend enough time with the assignments and provide the required number of comments, they receive full credit regardless of the quality of their engagement. In practice, we have found that most students engage meaningfully, but inevitably there are some for whom the Perusall comments become a box to be checked, and more broadly the grade is not meaningfully reflective of progress towards formational goals.

A potential downside to the low-friction implementation is that it can become easy in the busyness of a semester to let Perusall assignments run on autopilot in the background and fail to adequately integrate them into the rest of the course. This has, on occasion, led some students to perceive the readings as busywork. To counteract this perception, we have found it is important to regularly re-articulate the purpose of the readings and the broader formational goals of stewardship of information throughout the semester. When possible, we try to respond to or upvote student annotations as regularly as possible to demonstrate to students that we are actually reading and engaging with their ideas. Even when constraints on our time prevent in-depth engagement in Perusall itself, we try to still express appreciation for their engagement in class. Reading students thoughts and questions on Perusall readings is genuinely one of our favorite parts of our job, and we remind students of this regularly.

Because grades remain a significant motivating force for many students and because percentage of grade allocation is often interpreted as a measure of importance or time-worthiness, it would be beneficial to allocate a greater portion of a course grade to these formational learning goals captured by stewardship of information. However, lack of robust assessment strategies and reliance on completion grading remains a barrier. More work is needed to develop assessment strategies that can more substantively assess more subjective topics such as data ethics or civic engagement and can capture student formation over time.

Finally, having students engage with a variety of perspectives as well as socially and ethically complex ideas requires particular care. In addition to the examples given in Section 4 for creating an inclusive learning environment, we also recommend both keeping a close eye on student comments in Perusall as well as incorporating regular opportunities for students to give anonymous feedback (e.g., via surveys). This provides a safeguard for instructors to overcome some of the limitations of their own perspective and know when adjustment is needed to foster a more equitable learning environment conducive to the formation of all students.

## 6. **Conclusion**

In this paper, we assert that statistics and data science education are fundamentally an opportunity to develop students as good stewards of information. We provide both a conceptual framework and concrete examples for how to integrate this theme throughout a statistics and data science curriculum to foster the personal, professional, civic, and ethical formation of our students. The examples provided here suggest that relatively low-barrier strategies, such as curated weekly readings and structured reflections, can be a powerful and flexible tool for advancing these formational goals, and that implementation can be tailored to course content and audience in such a way that increases sophistication throughout the curriculum, including in introductory, advanced, applied, and theoretical courses.

Importantly, the ideas and examples presented here are intended to be illustrative not prescriptive. Our goal is not for all instructors to adopt the language of stewardship of information, though we have found it to be useful. Rather, we aim to provide a unifying lens that brings together many of the values and learning goals already present within statistics and data science education. We believe such a framework offers a way to integrate data literacy, evidence-based decision-making, communication, ethics, responsible use of data, and civic responsibility such that they become more than the sum of their parts. It creates space for an emphasis on statistics and data science as human endeavors and the fostering of an intellectual humility and principled approach to uncertainty that extends beyond the statistics classroom. Ultimately, stewardship of information invites us to move beyond only what students should know and towards who they are becoming. We hope to invite further imagination within the statistics and data science education for the types of statisticians, data scientists, citizens, and human beings we want to develop in our classrooms.